\documentclass[a4paper, 12pt]{article}

\usepackage[sort&compress]{natbib}
\bibpunct{(}{)}{;}{a}{}{,} 

\usepackage{amsthm, amsmath, amssymb, mathrsfs, multirow, url, subfigure}
\usepackage{graphicx} 
\usepackage{ifthen} 
\usepackage{amsfonts}
\usepackage[usenames]{color}
\usepackage{fullpage}

\theoremstyle{plain} 
\newtheorem{thm}{Theorem}

\theoremstyle{definition}

\theoremstyle{remark} 
\newtheorem*{pralg}{Predictive Recursion Algorithm}

\newtheorem{problem}{Problem}

\newcommand{\unif}{{\sf Unif}}
\newcommand{\nm}{{\sf N}}

\newcommand{\bin}{{\sf Bin}}

\newcommand{\RR}{\mathbb{R}}

\newcommand{\YY}{\mathbb{Y}}
\newcommand{\UU}{\mathbb{U}}

\renewcommand{\phi}{\varphi} 
\newcommand{\eps}{\varepsilon}

\newcommand{\fdr}{\mathrm{fdr}}
\newcommand{\fdrhat}{\widehat{\mathrm{fdr}}}

\title{On nonparametric estimation of a mixing density via the predictive recursion algorithm\footnote{This paper is dedicated to the memory of Professor Jayanta K.~Ghosh.}}
\author{Ryan Martin\footnote{Department of Statistics, North Carolina State University, {\tt rgmarti3@ncsu.edu}}}
\date{\today}

\begin{document}

\maketitle 

\begin{abstract}
Nonparametric estimation of a mixing density based on observations from the corresponding mixture is a challenging statistical problem.  This paper surveys the literature on a fast, recursive estimator based on the {\em predictive recursion} algorithm.  After introducing the algorithm and giving a few examples, I summarize the available asymptotic convergence theory, describe an important semiparametric extension, and highlight two interesting applications.  I conclude with a discussion of several recent developments in this area and some open problems.   

\smallskip

\end{abstract}

\section{Introduction}
\label{S:intro}

Estimating a mixing distribution based on samples from a mixture is arguably one of the most difficult statistical problems.  It boils down to estimating the distribution of a variable based on only indirect or noise-corrupted observations.  Nonparametric density estimation is already sufficiently challenging when one has direct observations let alone with only indirect observations.  But understanding this latent variable distribution has many important practical consequences so, despite the problem's difficulty, there are now a number of different methods available for estimating that distribution.  Here I will focus on a particular method, known as {\em predictive recursion} (PR), that provides a fast and easy-to-compute nonparametric estimate of a mixing density.  

The work on computation for Bayesian nonparametrics---in particular, for the Dirichlet process mixture model---in the late 1990s and early 2000s provided the original impetus for the development of PR.  At that time, Markov chain Monte Carlo (MCMC) for fitting Dirichlet process mixture models was an active area of research, e.g., \citet{escobar1994}, \citet{escobar.west.1995}, \citet{maceachern1994, maceachern1998}, \citet{maceachern.muller.1998}, and \citet{neal2000}, but computational power then was nowhere close to what it is now, so there was also an interest in developing alternatives to MCMC which were faster and easier in some sense.  At that time, Michael Newton and collaborators, in a series of papers \citep{nqz, newtonzhang, newton02}, developed the predictive recursion algorithm which aimed at providing a fast, MCMC-free approximation of the posterior mean of the mixing distribution under a Dirichlet process mixture model.  There was no doubt that the algorithm was fast and produced high-quality estimates in real- and simulated-data examples, but by the mid-2000s it was still unclear what specifically the PR algorithm was doing and what kind of properties the resulting PR estimator had.  Jayanta K.~Ghosh, or JKG for short, learned of the challenging open questions surrounding the PR algorithm and, naturally, was intrigued.  In 2005, he and his then student, Surya Tokdar, published the first fully rigorous investigation of the convergence properties of the PR estimator \citep{ghoshtokdar}.  Around that time, I was a PhD student at Purdue University looking for an advisor and a research project.  JKG generously shared with me a number of very promising ideas, but the one that stuck---and eventually became the topic of my thesis \citep{martinthesis}---was a deeper theoretical and practical investigation into the rather elusive PR algorithm.  

Between 2007 and 2012, JKG, Surya, and I were actively working on theory for and methodology based on PR.  The three of us eventually shifted our respective research foci to other things, but the developments continued.  In particular, James Scott and his collaborators found that PR is a powerful tool for handling the massive data and associated large-scale multiple testing problems arising in real-world applications.  I have also recently started working on some new PR-adjacent projects and those results shed light on the PR algorithm itself.  More on these efforts below.  

Sadly, on September 30th, 2017, JKG passed away, leaving a gaping hole in the scientific community that had once been overflowing with kindness and ingenuity.  Aside from his tremendous scholarly impact, JKG also touched the lives of many in a personal way.  I had the privilege of participating in several special JKG memorial conference sessions 
and I was moved by the many fond memories of JKG shared by the participants.\footnote{Anirban DasGupta's ``Remembering Professor Jayanta K.~Ghosh'' is an absolute must-read; see \url{http://www.stat.purdue.edu/news/2017/jayanta-ghosh.html}.}  To me, JKG was the epitome of a scientist: his research efforts were fueled by nothing other than an intense curiosity about the world, and his generosity as a teacher and mentor stemmed from an equally intense desire to share all that he knew.  


At face value, the goal of this paper is to review the PR algorithm, its theoretical properties, applications, and various extensions.  In particular, after a review of mixture models in Section~\ref{S:background}, I proceed in Section~\ref{S:pr} to define PR, give some illustrative examples, and summarize the basic theoretical convergence properties.  An important extension of PR is presented in Section~\ref{S:extension}, one that sets the scene for the applications described in Section~\ref{S:applications}.  At a higher level, however, the goal of this paper is to highlight an interesting albeit lesser-known area of statistics in which JKG had a major influence.  With this in mind, I present some recent developments and open problems in Sections~\ref{S:recent} and \ref{S:open}, respectively, in hopes of stimulating new research activity in this area and furthering JKG's legacy.  Section~\ref{S:discuss} gives some concluding remarks.

\section{Background on mixture models}
\label{S:background}

Consider independent and identically distributed (iid) data $Y_1,\ldots,Y_n$ with common density function given by the mixture model
\begin{equation}
\label{eq:mixture}
f(y) = f_p(y) = \int_\UU k(y \mid u) \, p(u) \, \nu(du), \quad y \in \YY \subseteq \RR^d. 
\end{equation}
Here $k(\cdot \mid u)$ is (for now) a fully known kernel, i.e., a density function with respect to, say, Lebesgue measure on $\YY$ for each $u \in \UU$, and $p$ is an unknown density with respect the given measure $\nu$ on $\UU$.  The goal is estimation of the {\em mixing} density $p$ based on iid data $Y_1,\ldots,Y_n$ from the {\em mixture} density $f$.  I will assume throughout that $p$ is identifiable, but this is non-trivial; see \citet{teicher61, teicher63} and \citet{quintana2002}.  Deconvolution is a special case of location mixtures, where $k(y \mid u) = k(y - u)$, and special techniques are available for this problem \citep{stefanski1990, zhang1990, zhang1995, fan1991}.  Here I will focus on methods for general mixture models.  

There are a number of approaches to this problem.  One is to give $p$ some additional structure, for example, to express $p$ as a discrete distribution.  This makes $f$ in \eqref{eq:mixture} a {\em finite mixture model} and producing maximum likelihood estimates (MLEs) of the parameters that characterize $p$, namely, the mixture weights and locations, can be readily found via, say, the EM algorithm \citep{dlr}.  One can alternatively give a prior distribution for the mixture weights and locations and then use, say, an EM-like data-augmentation strategy \citep[e.g.,][]{dyk.meng.2001} to sample from the posterior distribution and perform Bayesian inference.  


This approach, unfortunately, has some drawbacks.  In particular, the methods above can only be easily employed when the number of mixture components is known, which is an unrealistic assumption.  One can use model selection techniques, such as AIC \citep[e.g.,][]{leroux}, to select the number of components as part of a likelihood-based analysis.  Similarly, the Bayesian can put a prior distribution on the number of mixture components \citep[e.g.,][]{richardson}.  Ideally, one could let the data automatically choose the number of components, and there are nonparametric methods that can handle this.  Neither the nonparametric MLE \citep[e.g.,][]{lindsay1995, laird} nor the Dirichlet process mixture model \citep[e.g.,][]{muller.quintana.2004, ghosal2010} require the user to choose the number of mixture components.  In fact, JKG frequently worked with Dirichlet process mixture models; see \citet{ggr1999} and \citet{ghoshramamoorthi}.  

What makes estimation of $p$ difficult is that there are many different $p$ for which the corresponding mixture closely approximates the empirical distribution of $Y_1,\ldots,Y_n$.  That is, even if $p$ is identifiable, it is ``just barely so.''  Since the above methods are primarily focused on finding a $p$ such that the mixture \eqref{eq:mixture} fits the data well, there is no guarantee that the resulting $\hat p$ is a good estimate of $p$.  In fact, the nonparametric MLE is {\em discrete} almost surely \citep[][Theorem~21]{lindsay1995}, and the posterior mean of $p$ under a Dirichlet process mixture model also has some discrete-like features \citep[e.g.,][Figs.~1--2]{tmg}.  Therefore, if $p$ is assumed to be a smooth density, then a discrete estimator would clearly be unsatisfactory.  Smoothing of, say, the nonparametric MLE has been considered, but I will not discuss this here; see \citet{eggermont1995}.  One could also consider maximizing a penalized likelihood, one that encourages smoothness \citep{liulevinezhu2009}, but the computations are highly non-trivial.  

The mixture \eqref{eq:mixture} and the desire to estimate the mixing density manifests naturally when the model is expressed hierarchically.  That is, if unobservable latent variables $U_1,\ldots,U_n$ are iid $p$ and the conditional distribution of $Y_i$, given $U_i=u$, is $k(y \mid u)$, then the marginal distribution of $Y_i$ has a density of the form \eqref{eq:mixture}.  Often, the latent variables are the relevant quantities, e.g., measures of students' ``ability,'' so estimating $p$ would be of immediate practical interest.  This is a hopeless endeavor with only a few indirect observations from $p$, but, in the early 2000s, DNA microarray technologies changed this.  As \citet{efron2003} explains, this technology created a plethora of real-life problems where the individual $Y_i$ carries minimal information about its corresponding $U_i$ but the collection $(Y_1,\ldots,Y_n)$ carries a lot of information about $p$.  One way to take advantage of this information is to model $U_1,\ldots,U_n$ as {\em exchangeable} rather than iid, which amounts to assuming that the cases are ``similar'' in some sense.  This similarity suggests that it may be beneficial to share information across cases and, mathematically, the exchangeability assumption results in inference about $U_i$ that depend on all the data, not just on $Y_i$.  This type of ``borrowing strength'' \citep[e.g.,][p.~257]{ghosh-etal-book} was a central theme that emerged in much of JKG's later work, including \citet{bcfg2010, bogdan.ghosh.tokdar.2008}, \citet{dutta.bogdan.ghosh.2012}, and \citet{datta.ghosh.2013}.   An attractive alternative to a full hierarchical model, one that retains its ``borrowing strength'' feature, is an {\em empirical Bayes} solution, \`a la \citet{robbins1956, robbins1964, robbins1983}, where the data is used to estimate $p$.

\section{Predictive recursion}
\label{S:pr}

\subsection{Algorithm}

The methods described above are all likelihood-based, i.e., either the likelihood is optimized to produce an estimator or the likelihood is used to update a prior via Bayes's theorem, leading to a posterior distribution.  The predictive recursion (PR) algorithm, on the other hand, is not likelihood-based, at least not in its formulation.  Instead, PR processes the data points one at a time, using the following fast recursive update.  

\begin{pralg}
Initialize the algorithm with a guess $p_0$ of the mixing density and a sequence $\{w_i: i \geq 1\} \subset (0,1)$ of weights.  Given the data sequence $Y_1,\ldots,Y_n$ from the mixture model \eqref{eq:mixture}, evaluate 
\begin{equation}
\label{eq:recursion}
p_i(u) = (1-w_i) \, p_{i-1}(u) + w_i \, \frac{k(Y_i \mid u) p_{i-1}(u)}{f_{i-1}(Y_i)}, \quad i=1,\ldots,n, 
\end{equation}
where $f_{i-1}(y) = \int k(y \mid u) p_{i-1}(u) \, \nu(du)$ is the mixture corresponding to $p_{i-1}$.  Return $p_n$ and $f_n=f_{p_n}$ as the final estimates.  
\end{pralg}

Motivation for the PR algorithm, as described in \citet{nqz}, came from the simple and well-known formula for the posterior mean of $p$, under a Dirichlet process mixture model, based on a single observation.  That is, if the mixing distribution is assigned a Dirichlet process prior, with precision parameter $\alpha > 0$ and base measure with density $p_0$, then the posterior mean has density 
\[ \frac{\alpha}{\alpha+1} \, p_0(u) + \frac{1}{\alpha+1} \, \frac{k(Y_1 \mid u) p_0(u)}{f_0(Y_i)}, \]
which corresponds to the PR update with $w_i = (\alpha + i)^{-1}$.  Therefore, PR is exact in the case of $n=1$; I refer to this as the {\em one-step correspondence}.  For $n \geq 1$, Newton's proposal is simply to apply the one-step correspondence in each iteration, hence the PR algorithm is very straightforward: the output from the previous iteration is treated like a prior in the next, and the update is just a weighted average of the ``prior'' and its corresponding posterior based on a single data point.  This is an intuitively very reasonable idea, easy to implement, and fast to compute.  

Next are several important-but-quick observations about the PR algorithm.
\begin{itemize}
\item PR can estimate a density with respect to any user-specified dominating measure.  That is, if $p_0$ is a density with respect to $\nu$, then so is $p_n$ for all $n$.  Contrast this to the discrete nonparametric MLE and the ``rough'' \citep[][e.g., Figure~1]{tmg} Dirichlet process mixture posterior mean.  Having control the dominating measure gives the PR algorithm some advantages in certain applications; see Section~\ref{S:applications}.  
\vspace{-2mm}
\item The weight sequence $(w_i)$ affects PR's practical performance.  Theory in Section~\ref{SS:pr.properties} gives some guidance about the choice of weights, and examples usually take $w_i = (c+i)^{-\gamma}$ for some constants $c > 0$ and $\gamma \in (\frac12, 1]$.  
\vspace{-2mm}
\item The PR algorithm takes the form of {\em stochastic approximation} \citep{robbinsmonro}, which is designed for root-finding under measurement error.  This connection between the two recursive algorithms, fleshed out in \citet{martinghosh}, throws light on on the PR algorithm works.  Convergence properties for PR can be derived from general results for stochastic approximation \citep[e.g.,][]{pr-finite}, but this is so far limited to finite mixture cases.
\vspace{-2mm} 
\item One potentially concerning observation about the PR algorithm is that the final estimate, $p_n$, depends on the order in which the data sequence is processed.  In other words, at least in the iid case, $p_n$ is not a function of the sufficient statistic and, therefore, is not a Bayes estimate.  This dependence on the order is relatively weak when $n$ is large, and can be effectively eliminated by averaging over permutations of the data sequence.  This permutation-averaged PR estimator is just a Rao--Blackwellized version of the original PR estimator \citep{tmg}.  
\end{itemize}

\subsection{Illustrations}
\label{SS:pr.illustrations} 

\subsubsection{Poisson mixture}
\label{SSS:thai}

Example~1.2 in \citet{bohning} presents data $Y_1,\ldots,Y_n$ on the number of illness spells for $n=602$ pre-school children in Thailand over a two-week period.  The relatively large number of children---120 in total---with no illness spells makes these data {\em zero-inflated} and, therefore, a Poisson model is not appropriate.  This suggests a Poisson mixture model and here I will fit such a model, nonparametrically, using the PR algorithm.  

\begin{figure}
\begin{center}
\subfigure[Mixing distributions]{\scalebox{0.6}{\includegraphics{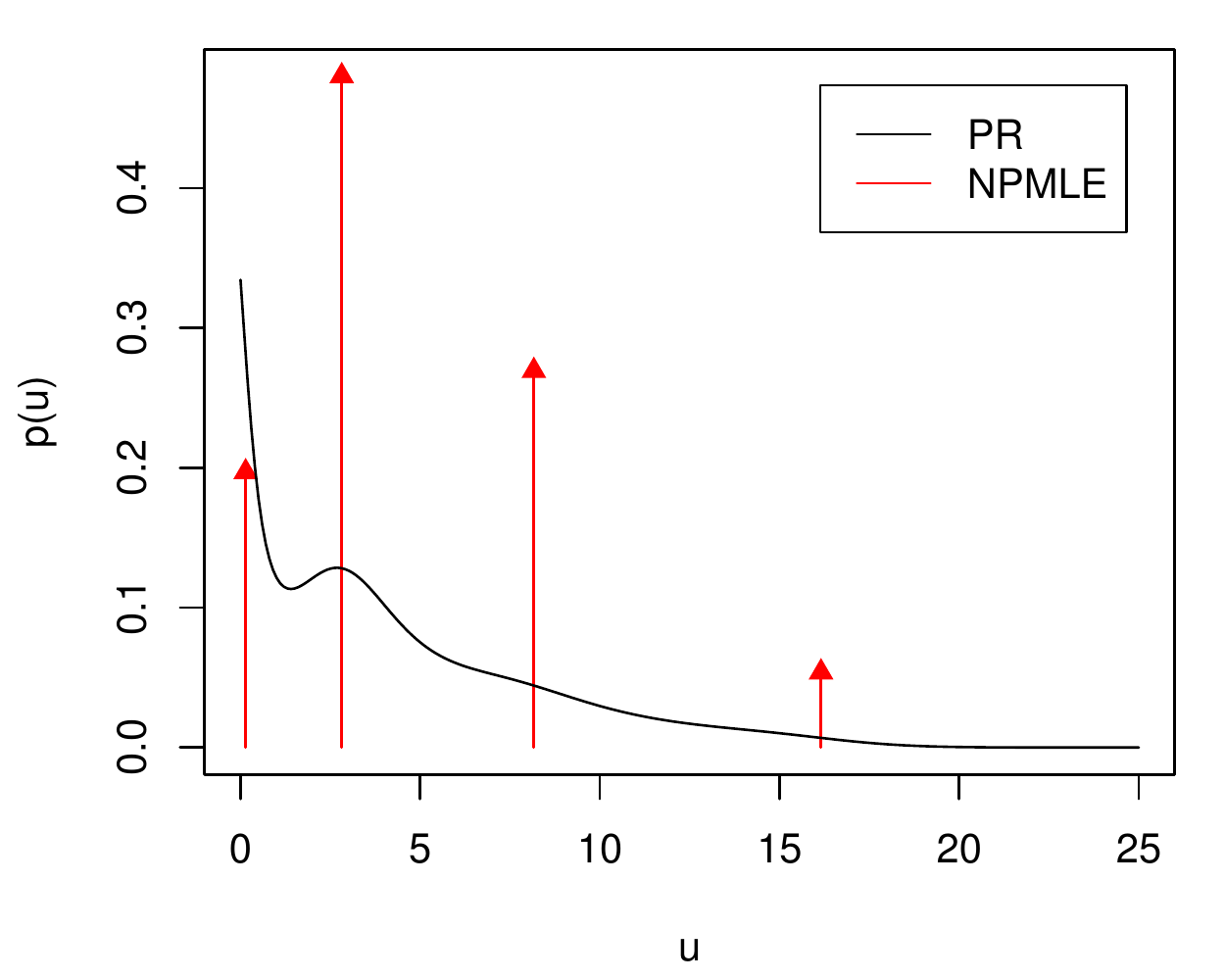}}}
\subfigure[Data and mixture distributions]{\scalebox{0.6}{\includegraphics{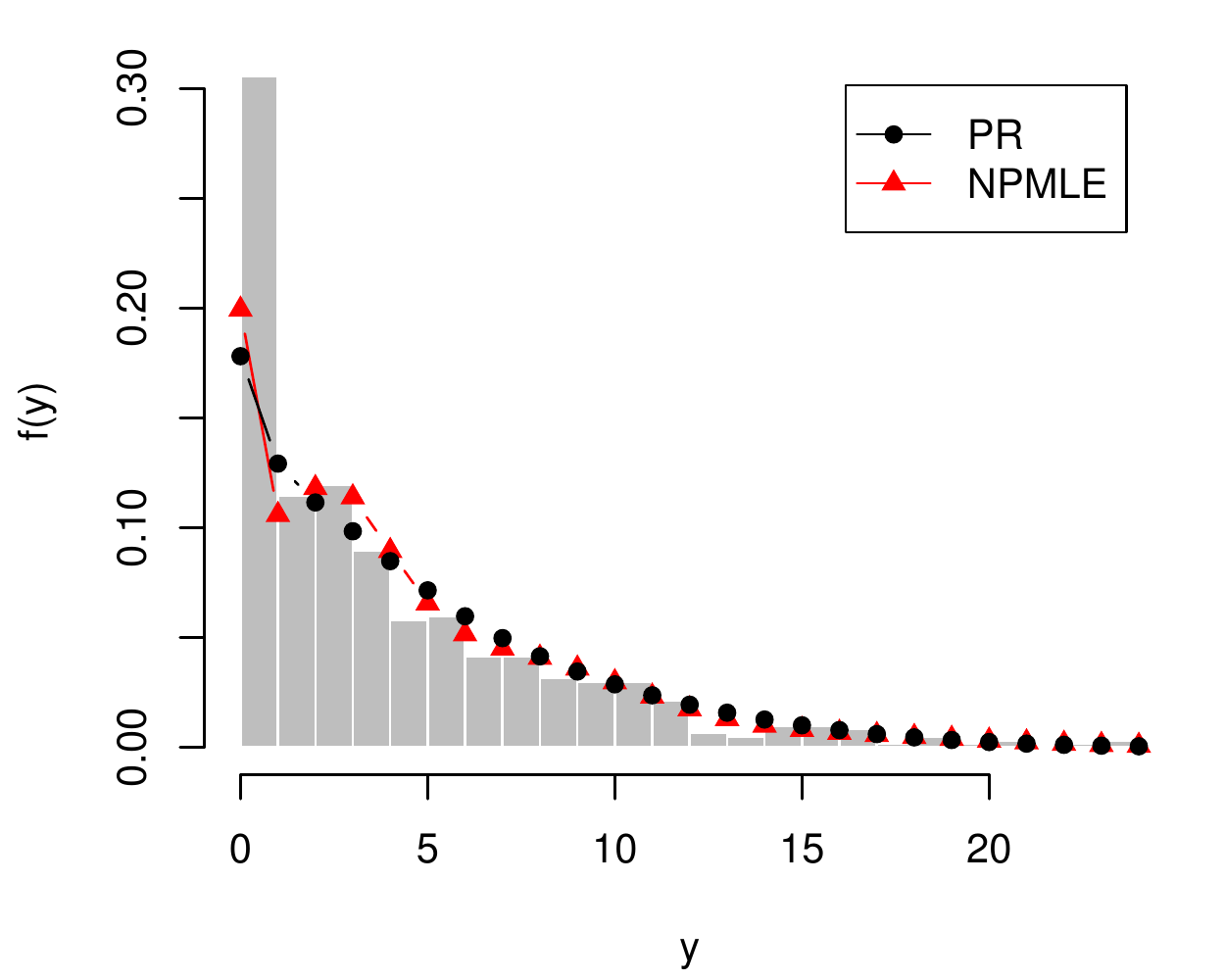}}}
\end{center}
\caption{Plots of the estimated mixing and mixture distributions---based on PR and nonparametric maximum likelihood---for the Poisson mixture example in Section~\ref{SSS:thai}.}
\label{fig:thai}
\end{figure}

In the mixture model formulation, $k(y \mid u)$ denotes a Poisson mass function with rate $u$, and $U_i$ represents, say, a latent ``healthiness'' index for the $i^{\text{th}}$ child.  Panel~(a) in Figure~\ref{fig:thai} shows the PR estimate of this density based on a $\unif(0,25)$ initial guess, weights as described above with $\gamma=0.67$, and 25 random permutations of the data sequence.  The relatively high concentration near 0 is consistent with the zero-inflation seen in the data.  Also shown in this panel is the nonparametric MLE, a discrete distribution, as presented in \citet[][Table~1]{wang}.  Note that the bump in the PR estimate around $y=3$ is consistent with the large mass assigned near $u=3$ by the nonparametric MLE.  But while the estimated mixing distributions are dramatically different, the two corresponding mixture distributions in Panel~(b) look very similar and both provide a good fit to the data.  Interestingly, the likelihood ratio of PR versus the nonparametric MLE is 0.98, very close to 1.  Therefore, within the class of mixing {\em densities}, there is little room to improve upon the PR estimator in terms of its quality of fit to the data; see, also, \citet{chae.martin.walker.npmle} and Section~\ref{SS:npmle} below.  

\subsubsection{Gaussian mixture}
\label{SSS:galaxy}

Gaussian mixture models, where $k(y \mid u)$ is a normal density with mean $u$ and variance either fixed or estimated from data, are widely used models for density estimation, clustering, etc.  Following \citet{roeder} and many others, I will consider data on the velocities (in thousands of km/sec) of $n=82$ galaxies moving away from Earth.  

\begin{figure}
\begin{center}
\subfigure[Mixing distributions]{\scalebox{0.6}{\includegraphics{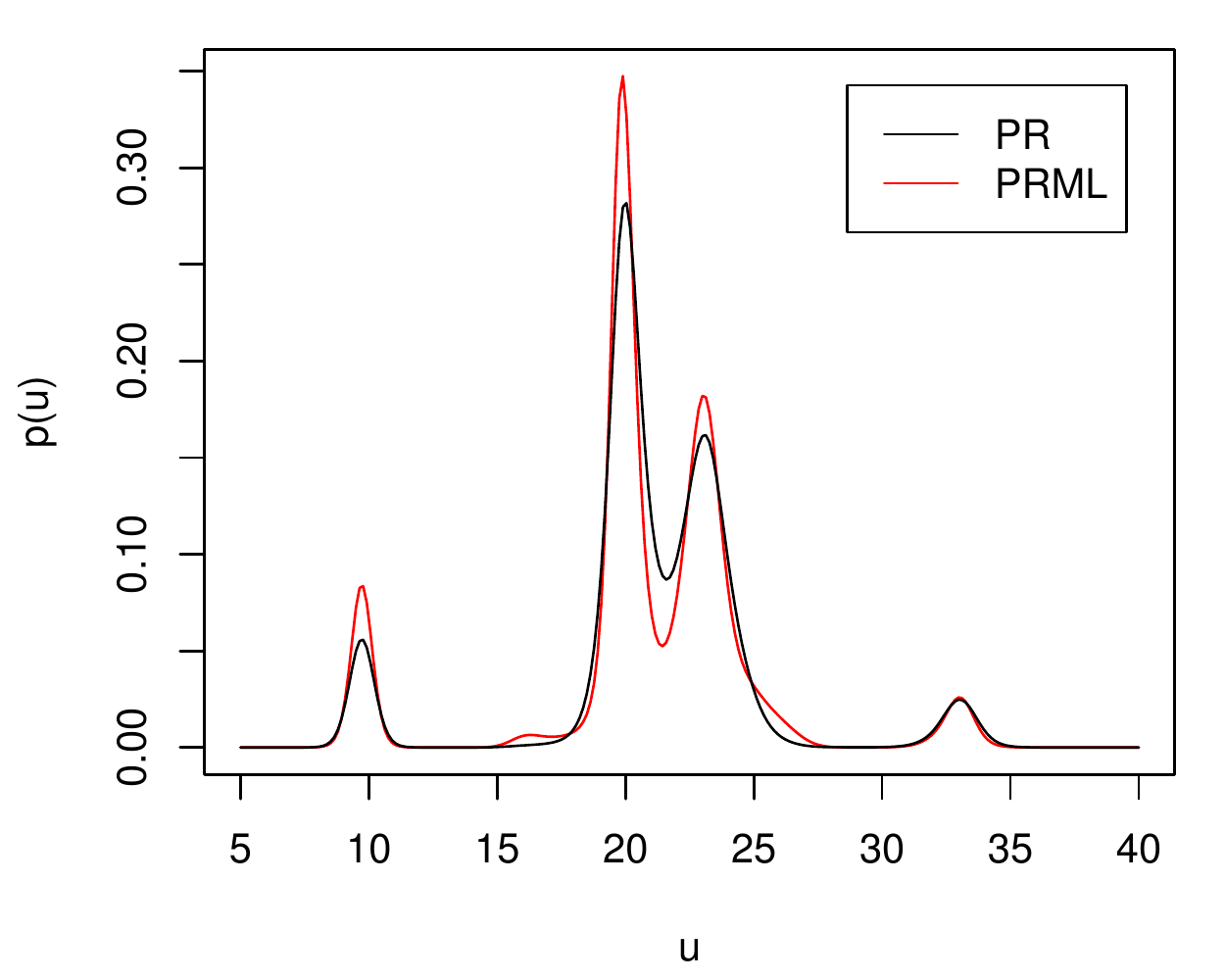}}}
\subfigure[Data and mixture distributions]{\scalebox{0.6}{\includegraphics{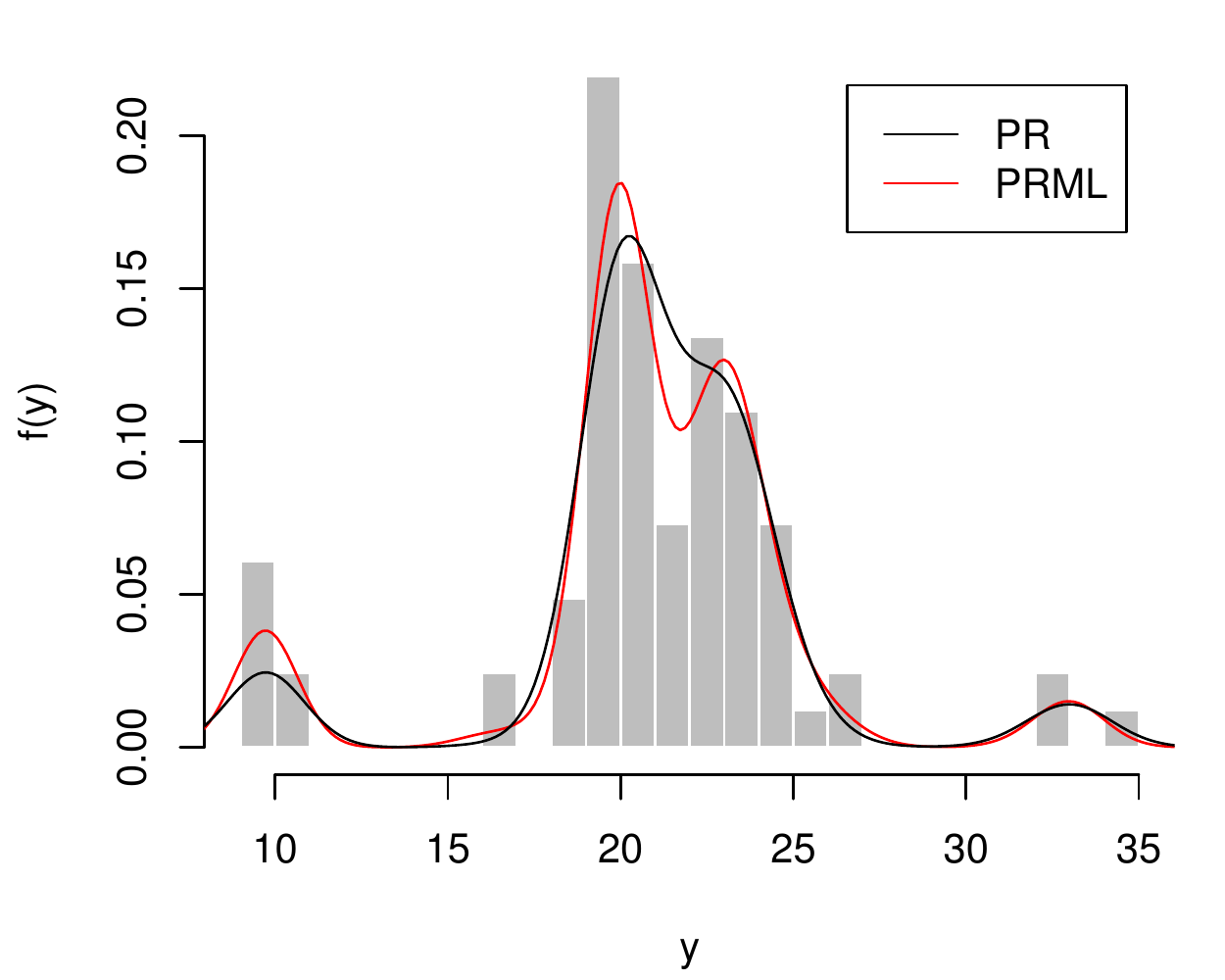}}}
\end{center}
\caption{Plots of the estimated mixing and mixture distributions---based on PR and its semiparametric extension, PRML, described in Section~\ref{S:extension}---for the Gaussian mixture example in Section~\ref{SSS:galaxy}.}
\label{fig:galaxy}
\end{figure}

Figure~\ref{fig:galaxy} shows the data histogram along with the PR estimates of the mixing and mixture distributions, in Panels~(a) and (b), respectively.  Here the PR algorithm uses the a kernel with standard deviation set at $\sigma=1$; the initial guess is $\unif(5,40)$ and the weights and permutation averaging is as in the previous example. The mixing density identifies four well-separated modes, but these are arguably not separated enough since the mixture appears to be a bit too smooth.  This is likely due to fixing the kernel scale parameter at $\sigma=1$.  The PR formulation can be extended naturally to semiparametric mixtures---see Section~\ref{S:extension}---and, here, I use this generalization to simultaneously estimate $p$ and the scale parameter $\sigma$.  The estimate in this case is $\hat\sigma=0.82$ and, as expected, the estimated mixing density has sharper peaks, leading to a less smooth and arguably better estimate of the mixture density.  


\subsubsection{Binomial mixture and empirical Bayes}
\label{SSS:nba}

In basketball, shots made from long distance count for 50\% more points than those from shorter distance.  These {\em three-point shots} can have a substantial effect on the outcome of a game, so three-point shooting performance strongly influences teams' offensive and defensive strategies.  I downloaded data from \url{www.nba.com} that lists the three-point shots made, $Y_i$, and attempted, $N_i$, for all $n=427$ NBA players in the last 10 games of the 2017--2018 season.  To study three-point shooting performance, I take $Y_i \sim \bin(N_i, U_i)$, independent, where $N_i$ is treated like a fixed covariate and $U_i$ represents the latent three-point shooting ability of player $i=1,\ldots,n$ during that crucial series of games at season's end.  Here I want to estimate the latent ability density, $p$, as part of an empirical Bayes analysis like in \citet{brown2008} and elsewhere for hitting in baseball.  

\begin{figure}
\begin{center}
\scalebox{0.7}{\includegraphics{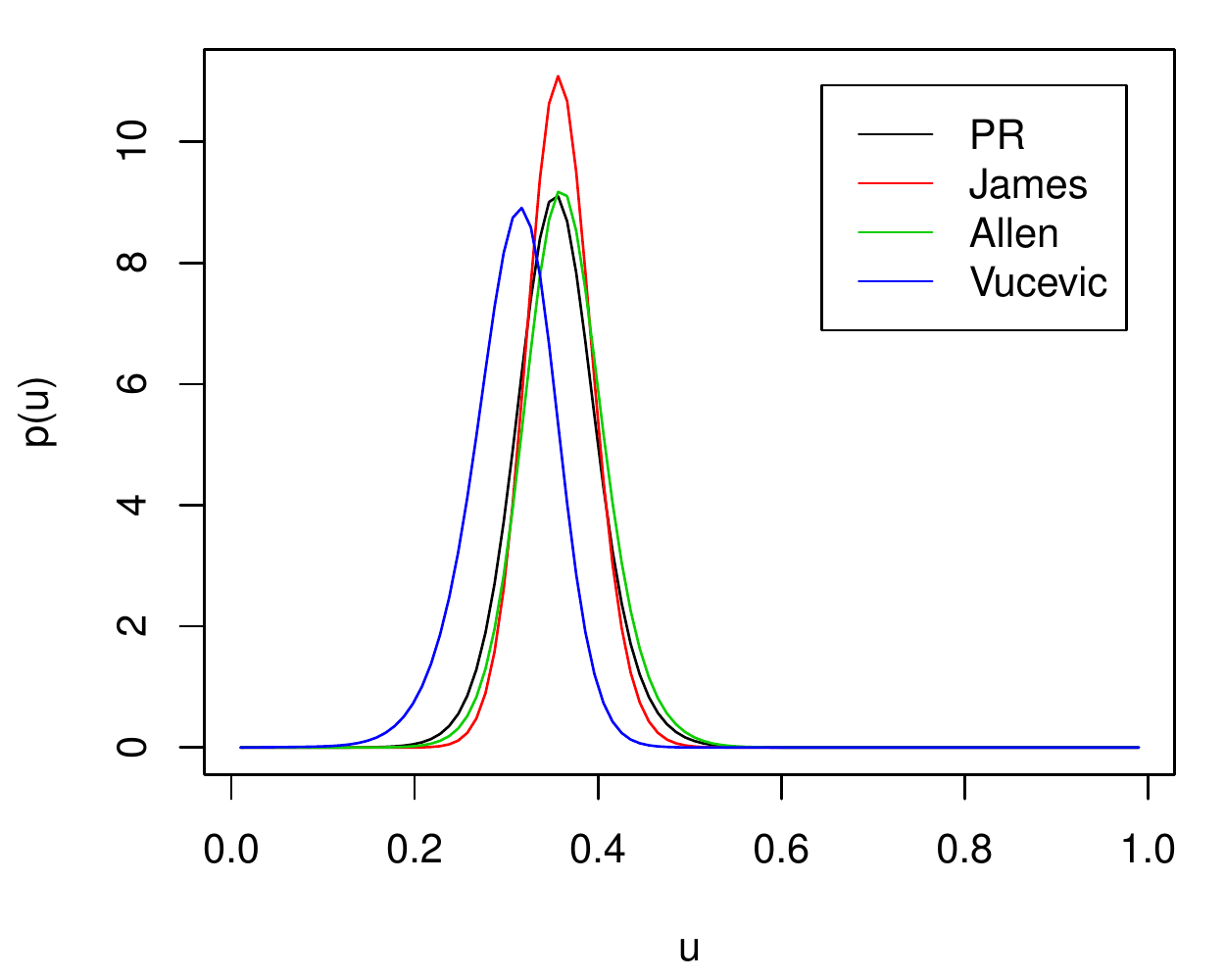}}
\end{center}
\caption{Plot of the PR estimate of the prior and the corresponding posterior for three selected NBA players in the three-point shooting example of Section~\ref{SSS:nba}.}
\label{fig:nba}
\end{figure}

The solid black line in Figure~\ref{fig:nba} shows the PR estimate of the prior density $p$ based on a $\unif(0,1)$ initial guess and weights and permutation averaging as in the previous examples.  This is unimodal, with mode 0.36, and concentrates about all its mass in the interval $(0.2, 0.6)$.  The other lines in the plot show the corresponding empirical Bayes posterior densities for three selected players, namely, LeBron James, Jarret Allen, and Nikola Vucevic, whose proportion of three-point shots made for this series of games was 19/52, 2/3, and 1/18, respectively.  James's proportion is very close to the estimated prior mode and his number of attempts is high, so his estimated posterior is a more-concentrated version of the prior.  Allen's proportion of makes is high compared to the prior mode, but the number of attempts is low, hence strong shrinkage towards the prior mode.  Finally, Vucevic's proportion is very low but based on a moderate number of attempts, so only a moderate amount of shrinkage towards the prior mode.

\subsection{Theoretical properties}
\label{SS:pr.properties}

Since the PR output $p_n$ is neither a maximum likelihood nor a Bayesian estimator, its convergence properties do not follow immediately from the standard asymptotic theory, so something different is needed.  \citet{ghoshtokdar} gave the first rigorous results on convergence of the PR estimator, using martingale techniques, which were later extended in \citet{tmg} and again in \citet{mt-rate}.  

As before, let $Y_1,\ldots,Y_n$ be iid samples from a density $f^\star$, but allow the possibility that the posited mixture model is {\em misspecified}, that is, the common marginal density $f^\star$ may not have a mixture representation as in \eqref{eq:mixture}.  In this misspecified case, since there may not be a ``true'' mixing density, it is not entirely clear what it means for the PR estimator to converge.  The best one could hope for is that the PR estimate, $f_n$, of the marginal density would converge to the ``best possible'' mixture of the specified form \eqref{eq:mixture}.  More specifically, if $K$ denotes the Kullback--Leibler divergence, then, ideally, $K(f^\star, f_n)$ would converge to $\inf_f K(f^\star, f)$, where the infimum is over the set of mixtures in \eqref{eq:mixture} for the given kernel, etc.  Conditions under which the infimum is attained for a mixture $f^\dagger=f_{p^\dagger}$, with corresponding mixing density $p^\dagger$, are given in, e.g., \citet[][Lemma~3.1]{mt-rate} and \citet[][Lemma~3.1]{kleijn}; recall that I assume the mixture model is identifiable, so this $p^\dagger$ is unique.  Of course, if the mixture model is well-specified, then $f^\dagger = f^\star$ and $p^\dagger$ equals the true mixing density, $p^\star$.  

Naturally, the PR convergence theorem requires some assumptions.  There are two sets of conditions, one on the posited mixture model and the other on the PR algorithm's inputs.  I briefly summarize each in turn.
\begin{itemize}
\item For the mixture model, more general results are available, but here I will assume that the mixing densities are all fully supported on a compact set $\UU$.  I will also assume that the kernel is such that $u \mapsto k(y \mid u)$ is bounded and continuous for almost all $y$.  Finally, certain integrability of density ratios is needed in the proof, so it will be assumed that 
\begin{equation}
\label{eq:integrable}
\sup_{u_1,u_2 \in \UU} \int \Bigl\{ \frac{k(y \mid u_1)}{k(y \mid u_2)} \Bigr\}^2 \, f^\star(y) \, dy < \infty. 
\end{equation}
This is a strong condition, but, since $\UU$ is assumed to be compact, it holds if $k(y \mid u)$ is an exponential family and $f^\star$ has Gaussian-like tails.  
\item For the PR algorithm's inputs, namely, the initial guess $p_0$ and the weight sequence $(w_i)$, the assumptions are quite mild.  First, it is necessary that the support of $p_0$ contain that of $p^\dagger$.  If the compact support $\UU$ is known, then this is trivially satisfied.  Second, the weights must satisfy 
\begin{equation}
\label{eq:weights}
\sum_{i=1}^\infty w_i = \infty \quad \text{and} \quad \sum_{i=1}^\infty w_i^2 < \infty. 
\end{equation}
The suggested class of weights, $w_i = (c + i)^{-\gamma}$, for $\gamma \in (\frac12, 1]$ satisfy this. 
\end{itemize}

The following theorem summarizes the known convergence properties of the PR estimators $p_n$ and $f_n$.  A version of the consistency result below, in the well-specified case, is also presented in Section~5.4 of \citet{ghosal.vaart.book}.  

\begin{thm}
\label{thm:pr}
Assume that $Y_1,Y_2,\ldots$ are iid samples from density $f^\star$ and that the aforementioned conditions are met.  Set $K_n = K(f^\star, f_n) - \inf_p K(f^\star, f_p)$.
\begin{enumerate}
\item Then $K_n \to 0$ almost surely.
\vspace{-2mm}
\item If $\sum_n a_n w_n^2 < \infty$, where $a_n = \sum_{i=1}^n w_i$, then $a_n K_n \to 0$ almost surely.  
\vspace{-2mm}
\item If the kernel is tight in the sense of \citet[][Condition~A6]{mt-rate}, then $p_n$ converges weakly to $p^\dagger$ almost surely.  
\end{enumerate}
\end{thm}

An interesting by-product of the proof of Theorem~\ref{thm:pr} creates an asymptotic link between PR and the nonparametric MLE.  That is, the PR estimator, $p_n$, is converging to a solution $P^\dagger$, which may or may not have a density, such that 
\[ \int \frac{k(y \mid u)}{f_{P^\dagger}(y)} \, f^\star(y) \, dy = 1 \quad \text{for $P^\dagger$-almost all $u$}. \]
But according to \citet[][p.~115]{lindsay1995}, the nonparametric maximum likelihood estimator $\hat P$ is characterized as a solution to 
\[ \frac1n \sum_{i=1}^n \frac{k(Y_i \mid u)}{f_{\hat P}(Y_i)} = 1 \quad \text{for $\hat P$-almost all $u$}. \]
When $n$ is large, the above average is approximately equal to the expectation with respect to $f^\star$, hence a link between the PR algorithm's target and the nonparametric MLE.  

The first and third claims in Theorem~\ref{thm:pr} establish consistency of the PR estimates.  The compactness condition eluded to in the third claim holds for all the standard kernels so it imposes no practical constraints.  

The second claim in Theorem~\ref{thm:pr} gives a bound on the PR rate of convergence.  That condition is satisfied for $w_i = (c+i)^{-\gamma}$ for $\gamma \in (\frac23,1]$, and gives a corresponding Kullback--Leibler convergence rate for $f_n$ of about $n^{-1/3}$.  Unfortunately, this leaves something to be desired.  For example, \citet{ghosalvaart2001} showed that, with a Gaussian kernel and a Dirichlet process prior on the mixing distribution, the Bayes posterior concentrates around a true Gaussian mixture at nearly a $n^{-1}$ rate in Kullback--Leibler divergence.  But the PR rate above makes no assumptions about the true density $f^\star$ so it is interesting to understand the nature of that rate.  \citet{mt-rate} showed that PR's $n^{-1/3}$ rate is ``minimax'' in nature, i.e., it is the rate PR attains when $f^\star(y) = k(y \mid u^\star)$ for some fixed $u^\star$ value, the ``most extreme'' kind of mixture where ``$p^\star$'' is a point mass at $u^\star$.  See Section~\ref{S:open}.


\section{Semiparametric mixture extension}
\label{S:extension}

So far, I have assumed that the kernel $k$ in the mixture model is fixed.  However, there are cases in which it would make sense to allow the kernel to depend on some other parameters, say, $\theta$, that do not get mixed over.  The standard example would be to allow a Gaussian kernel to depend on some scale parameter while being mixed over the mean; see below.  That is, here I am concerned with a semiparametric mixture model where the goal is to simultaneously estimate both the mixing density $p$ and the non-mixing structural parameter $\theta$.  For this, it turns out that the asymptotic theory for PR under model misspecification plays an important role.  

Write the $\theta$-dependent kernel as $k_\theta(y \mid u)$, and let $p_{i,\theta}$ denote the PR estimate of the mixing density based on $Y_1,\ldots,Y_i$, with kernel $k_\theta$ fixed throughout.  Also write $f_{i,\theta}(y) = \int k_\theta(y \mid u) \, p_{i,\theta}(u) \, \nu(du)$ for the corresponding mixture.  Next, define a sort of ``likelihood function'' based on the PR output, that is, 
\begin{equation}
\label{eq:prml}
L_n(\theta) = \prod_{i=1}^n f_{i-1,\theta}(Y_i).
\end{equation}
\citet{mt-prml} motivated this choice of likelihood by showing that $L_n(\theta)$ had features resembling that of the marginal likelihood for $\theta$ under a fully Bayesian Dirichlet process mixture model.  I will refer to \eqref{eq:prml} as the {\em PR marginal likelihood}, and I proceed to estimate the structural parameter by maximizing this function.  

For a quick example, consider a kernel $k_\theta(y \mid u) = \nm(y \mid u, \theta^2)$.  The likelihood function in \eqref{eq:prml} can be readily evaluated and maximized numerically to simultaneously estimate $p$ and $\theta$.  This approach was carried out in the galaxy data example of Section~\ref{SSS:galaxy} and the additional flexibility of being able to estimate the kernel scale parameter via PR marginal likelihood optimization resulted in an estimated mixture density that fit the data histogram better compared to that from the original PR.  

Maximizing $L_n(\theta)$ is equivalent to minimizing $n^{-1} \sum_{i=1}^n \log\{f^\star(Y_i)/f_{i-1,\theta}(Y_i)\}$, and it follows from Theorem~\ref{thm:pr} that this latter function converges pointwise, as $n \to \infty$, to $\inf_p K(f^\star, f_{p,\theta})$, where the infimum is over all mixing densities.  Therefore, at least intuitively, one would expect that 
\begin{equation}
\label{eq:consistency}
\hat\theta \to \arg\min_\theta \bigl\{ \inf_p K(f^\star, f_{p,\theta}) \bigr\}, \quad n \to \infty. 
\end{equation}
It turns out, however, that this consistency property is quite difficult to demonstrate in general; see Section~\ref{S:open}.  But numerical results in \citet{mt-prml} and elsewhere suggest that \eqref{eq:consistency} does hold and, moreover, so does asymptotic normality.

\section{Applications}
\label{S:applications}

There are a number of applications of the PR algorithm and its semiparametric extension in the literature.  See \citet{taonewton1999}, \citet{newtonzhang}, the example in \citet{newton02} based on the genetics application in \citet{newton-ecoli}, \citet{todem.williams.2009}, and the very recent work by \citet{woody.scott.post} on valid Bayesian post-selection inference.  Here I only highlight two specific applications, one in large-scale significance testing, an area in which JKG worked, and one in robust regression.

\subsection{Large-scale significance testing}
\label{SS:testing}

In the hierarchical model formulation at the end of Section~\ref{S:background}, consider a large collection $U_1,\ldots,U_n$ of latent variables where case $i$ is said to be ``null'' if $U_i=0$ and ``non-null'' otherwise.  An example is DNA microarray experiments where the cases correspond to genes and ``null'' means that the gene is not differentially expressed.  Of course, only noisy measurements $Y_1,\ldots,Y_n$ of $U_1,\ldots,U_n$ are available, so the goal is to test the sequence of hypotheses, $H_{0i}: U_i = 0$ versus $H_{1i}: U_i \neq 0$, $i=1,\ldots,n$.  What makes this an interesting statistical problem is that $n$ is large and most of the cases are null, e.g., most genes are not associated with a particular phenotype, so it is beneficial to share information across cases.  Brad Efron wrote extensively on empirical Bayes solutions this problem in the early 2000s \citep[e.g.,][]{efron2010book}, and here I will summarize a PR-based implementation of Efron's approach presented in \citet{mt-test}.  Recent extensions of this proposal to handle covariates and certain spatial dependence are presented in \citet{scott.FDRreg} and \citet{scott.fdrsmooth}, respectively.  

\citet{efron2008} describes the {\em two-groups model} where $Y_1,\ldots,Y_n$ are assumed to have a common density function of the form 
\begin{equation}
\label{eq:twogroups}
f(y) = \pi \, f^{(0)}(y) + (1-\pi) \, f^{(1)}(y),
\end{equation}
where $f^{(0)}$ and $f^{(1)}$ correspond to the densities under null and non-null settings, respectively, and $\pi$ represents the proportion of null cases.  He argues that, basically without generality, one can take $f^{(0)}(y) = \nm(y \mid \mu, \sigma^2)$, but perhaps with parameters $(\mu,\sigma^2)$ that need to be estimated, i.e., an {\em empirical null} \citep{efron2004}.  Assuming, for the moment, that all the pieces in \eqref{eq:twogroups} are known, one can show that the Bayes test of $H_{0i}$ would reject if $\fdr(Y_i) \leq c$, for $c=0.1$, say, where fdr---the {\em local false discovery rate}---is given by 
\begin{equation}
\label{eq:fdr}
\fdr(y) = \pi f^{(0)}(y) / f(y). 
\end{equation}
Efron's insight was that, since $n$ is large, nonparametric estimation of the marginal density is straightforward and, likewise, since most of the cases are null, $(\pi, \mu, \sigma^2)$ could also be estimated.  Plugging these estimates into the expression \eqref{eq:fdr} and carrying out the sequence of tests with the corresponding estimate of fdr is Efron's empirical Bayes solution.  Details can be found in, e.g., \citet{efron2004}, and alternative estimation strategies are given in \citet{jincai2007}, \citet{muralidharan2009}, \citet{jin.peng.wang.2010}, and \citet{jeng.zhang.tzeng.2018}.  

An advantage of Efron's approach is that it is apparently not necessary to directly model the possibly complicated non-null density $f^{(1)}$.  However, it is possible that the independent estimates of $f$ and $f^{(0)}$ are incompatible in the sense that, e.g., $\hat\pi \hat f^{(0)}(y) > \hat f(y)$ for some $y$.  To avoid such issues, a model for all the ingredients in \eqref{eq:twogroups}---one that is sufficiently flexible in $f^{(1)}$---is needed.  Toward this, \citet{mt-test} embed \eqref{eq:twogroups} into the general mixture formulation \eqref{eq:mixture} by taking the dominating measure 
\[ \nu(du) = \delta_0(du) + \lambda_{[-1,1]}(du), \]
a point-mass at 0 plus Lebesgue measure on $[-1,1]$, and kernel 
\[ k_\theta(y \mid u) = \nm(y \mid \mu + \tau \sigma u, \sigma^2), \quad \theta=(\mu, \tau, \sigma). \]
With these choices, the mixture in \eqref{eq:mixture} takes the form
\begin{equation}
\label{eq:mt.twogroups}
f(y) = \pi \nm(y \mid \mu, \sigma^2) + (1-\pi) \int_{-1}^1 \nm(y \mid \mu +  \tau\sigma u, \sigma^2) \, p(u) \, du, 
\end{equation}
which can immediately be identified as a model-based version of \eqref{eq:twogroups}.  Intuitively, the non-null cases, which correspond to ``signals,'' should tend to be larger magnitude, so it makes sense that $f^{(1)}$ have heavier tails than $f^{(0)}$.  The normal location mixture in \eqref{eq:mt.twogroups} can achieve this, and the parameter $\tau$ controls roughly how much heavier the normal the tails need to be.  Since the PR algorithm respects the specified dominating measure, the combined discrete-continuous form of the mixing distribution can be handled easily, and (a minor modification of) the semiparametric extension of PR in Section~\ref{S:extension} can be applied to fit the model in \eqref{eq:mt.twogroups} and define the corresponding empirical Bayes testing procedure based on the plug-in estimate of fdr.  

\begin{figure}
\begin{center}
\scalebox{0.7}{\includegraphics{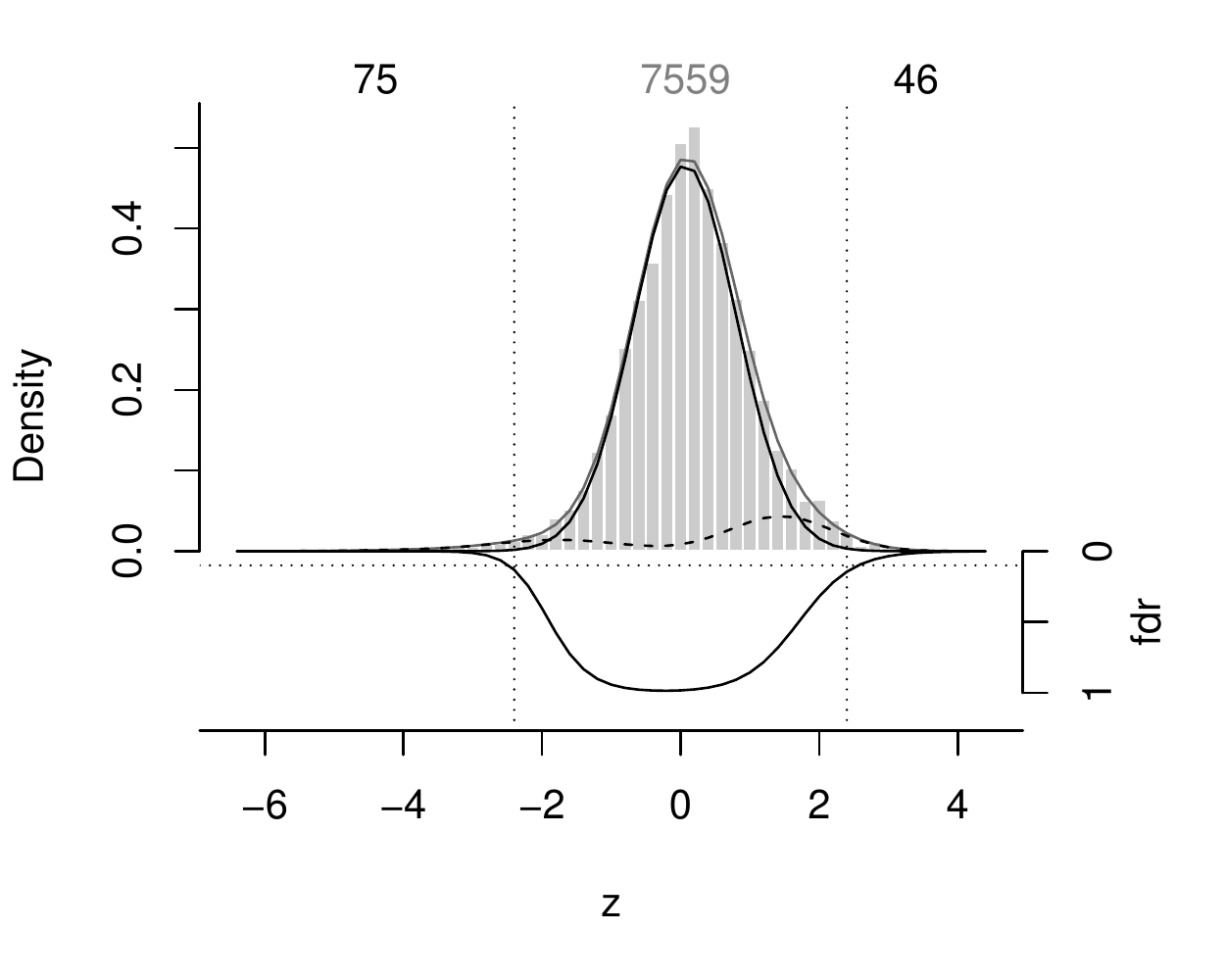}}
\end{center}
\caption{Histogram of the z-scores for the HIV data in \citet{vant2003}, along with estimates of $f$, $\pi f^{(0)}$, and $(1-\pi) f^{(1)}$ overlaid.  The inverted scale shows the estimated fdr and the $\fdrhat \leq 0.1$ cutoffs for the test.}
\label{fig:hiv}
\end{figure}

For illustration, I consider data from the study in \citet{vant2003} that compares the genetic profiles of four healthy and four HIV-positive patients.  The goal is to determine which, if any, of the $n=7680$ genes are differentially expressed between the two groups.  This example is described in \citet[][Section~6.1D]{efron2010book}.  Figure~\ref{fig:hiv} shows the results of the PR model fit; in particular, $\hat\mu=0.07$, $\hat\sigma=0.74$, and $\hat\pi=0.88$.  The estimated $f$ clearly fits the data histogram, which is wide enough to leave room for the normal $f^{(0)}$ and the heavier-tailed bimodal estimate of $f^{(1)}$.  The inverted scale shows the estimated fdr and the ``$\fdrhat \leq 0.1$'' cutoffs are also show.  Finally, the plot indicates that 121 genes are identified by the test as differentially expressed, 46 are up- and 75 are down-regulated.  The conclusions here are similar to those obtained by Efron, but this is not always the case; cf.~\citet{mt-test}.  


\subsection{Robust regression}

Consider a linear regression model where 
\[ y_i = x_i^\top \beta + \eps_i, \quad i=1,\ldots,n, \]
where $y_i$ is a real-valued response, $x_i$ is a $d$-vector of predictor variables, $\beta$ is a $d$-vector of regression coefficients, and $\eps_i$ are measurement errors, assumed to be iid.  Quantification of uncertainty about estimates or predictions in this setting requires specification of a distribution for the errors.  A standard choice is to assume the errors are normal, leading to simple closed-form expressions for the MLEs with straightforward sampling distribution properties.  However, if the normal error assumption is questionable, e.g., if there are ``outliers,'' then the MLEs will suffer.  Therefore, it is of interest to develop procedures that can handle different error distribution assumptions, especially those with heavier-than-normal tails.  It is indeed possible to introduce a heavy-tailed error distribution, such as Student-t with small degrees of freedom, and work out the corresponding MLEs and their properties, but this is still an assumption that may not be appropriate for the given problem.  A more flexible, nonparametric choice of error distribution would desirable.  Motivated by the fact that scale mixtures of normals produce heavy-tailed distributions, \citet{prreg} introduce a mixture model formulation and propose to estimate both the mixing density and $\beta$ using the semiparametric extension of PR described in Section~\ref{S:extension}.  Here I briefly summarize their approach.

With a slight abuse of my previous notation, let me write $f$ for the density function of the measurement errors, $\eps_1,\ldots,\eps_n$.  Expressing $f$ as the mixture 
\[ f(\eps) = \int_0^\infty \nm(\eps \mid 0, u^2) \, p(u) \, du, \]
for some unknown mixing density $p$, is one way to induce a flexible, heavy-tailed distribution for the errors.  For any fixed $\beta$, by writing $\eps_i = y_i - x_i^\top \beta$, the PR algorithm can be used to estimate the mixing and mixture densities, $p$ and $f$, respectively.  Of course, those estimates would depend on $\beta$ so, like in Section~\ref{S:extension}, I could define a marginal likelihood in $\beta$ to be maximized, leading to a simultaneous estimate of $\beta$ and $p$.  Optimization of this marginal likelihood is non-trivial, but \citet{prreg} propose a hybrid PR--EM algorithm wherein they introduce latent variables $U_i$ from the mixing distribution to make the ``complete-data'' likelihood of a simple Gaussian form.  Details are in their paper and an R code implementation is available at my website.  \citet{sriboonchitta.prml} used a similar PR--EM strategy in a time series application.

\begin{figure}
\begin{center}
\scalebox{0.7}{\includegraphics{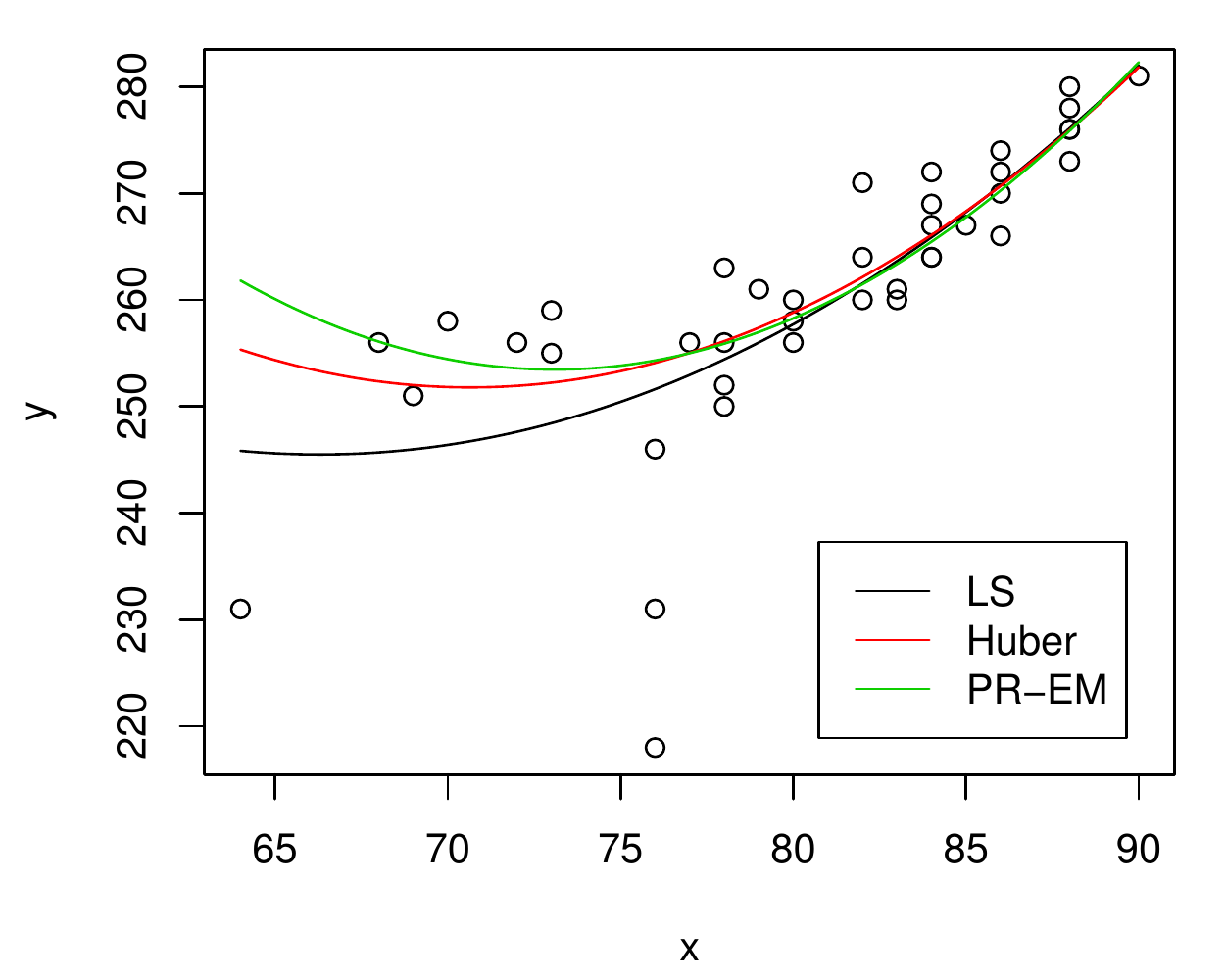}}
\end{center}
\caption{Scatter plot of the mathematics proficiency score data and the three different quadratic model fits.}
\label{fig:reg}
\end{figure}

As an example, I consider data on mathematics proficiency presented in Table~11.4 of \citet{kutner.etal.book}.  The response variable, $y$, is the students' average mathematics proficiency exam score for 37 U.S.~states, the District of Columbia, Guam, and the Virgin Islands; hence, $n=40$.  The predictor variable, $x$, is the percentage of students in each state with at least three types of reading materials at home.  This is an interesting example because D.C.~and Virgin Islands are outliers in $y$ and Guam is an outlier in both $x$ and $y$.  The general trend suggests a quadratic model, 
\[ y_i = \beta_0 + \beta_1 x_i + \beta_2 x_i^2 + \eps_i, \quad i=1,\ldots,n, \]
and the plot in Figure~\ref{fig:reg} shows the data and the results of three fits of the above model, namely, ordinary least squares, Huber's robust least squares, and PR--EM.  Here the former two methods are both more influenced by the outliers than the PR--EM method, suggesting that the latter puts lesser weight on those extremes in the model fit.

\section{Recent developments}
\label{S:recent}

\subsection{PR for the mixture}

The PR algorithm is designed for estimating the mixing distribution but, of course, it is at least conceptually straightforward to produce a corresponding estimate for the mixture distribution.  However, the PR algorithm requires numerical integration at each iteration, which itself requires that the mixing density support be known, compact, and no more than two dimensions.  If the sole purpose of the mixture model was to facilitate density estimation, as is often the case, then the above requirements are a hindrance.  It is, therefore, natural to ask if it is possible to formulate a numerical integration-free version of the PR algorithm directly on the mixture density.  \citet{hahn.martin.walker.pred} happened upon an affirmative answer to this question while investigating a seemingly unrelated updating property of Bayesian predictive distributions.  

Given a prior density $p_0(u)$ and a kernel $k(y \mid u)$, let $f_0$ be the prior predictive density, with corresponding distribution function $F_0$.  For a sequence of data $Y_1,Y_2,\ldots$, let $f_i$ denote the Bayesian posterior predictive distribution for $Y_{i+1}$, given $Y_1,\ldots,Y_i$.  Then \citet{hahn.martin.walker.pred} showed that there exists a sequence of bivariate copula densities $c_i$ such that $f_i(y) = c_i(F_{i-1}(y), F_{i-1}(Y_i)) f_{i-1}(y)$.  That is, if this sequence of copula densities were known, then one could recursively update the Bayesian predictive distribution without any posterior sampling, MCMC, etc.  For simple Bayesian models, the closed-form expressions for the copula densities can be derived, but not in general.  

Indeed, for a Dirichlet process mixture model, only the first in the sequence of copula densities can be derived in closed-form.  This suggests following Newton's strategy, capitalizing on the one-step exactness of the recursive update, to derive a new algorithm.  The specific proposal in \citet{hahn.martin.walker.pred} is the update 
\[ f_n(y) = (1-w_n) f_{n-1}(y) + w_n g_\rho( F_{n-1}(y), F_{n-1}(Y_n)) \, f_{n-1}(y), \]
where $g_\rho$ is the Gaussian copula density with correlation parameter $\rho$.  Those authors show that this algorithm is fast to compute and provides accurate estimate in finite-sample simulation experiments.  They also prove consistency under tail conditions on the true density.  It would be interesting to investigate convergence rates and to extend this method to handle multivariate and dependent data sequences.

\subsection{A variation on PR}
\label{SS:npmle}

A potentially troubling feature of the PR algorithm is its dependence on the order of the data sequence.  Averaging over permutations reduces this dependence, but is not a fully satisfactory fix. Therefore, other similar algorithms might be of interest.  

One that has made an appearance in numerous places across the literature, but has yet to be systematically studied, is as follows.  Start by assuming that the mixture density $f$ in \eqref{eq:mixture} is {\em known}.  Then the algorithm 
\begin{equation}
\label{eq:fredholm}
p_t(u) = \int \frac{k(y \mid u) f(y)}{\int k(y \mid v) \, p_{t-1}(v) \,dv} \, dy, \quad t \geq 1, 
\end{equation}
will converge to a solution of the inverse problem defined in \eqref{eq:mixture}; see \citet{chae.martin.walker.fredholm}.  In a statistical context, where $f$ is unknown but data $Y_1,\ldots,Y_n$ is available, there are a number of ways one can modify the above algorithm.  One is to replace $f$ in \eqref{eq:fredholm} with the empirical distribution of $Y_1,\ldots,Y_n$.  This produces a smooth mixing density estimate at every finite $t$, but \citet{chae.martin.walker.npmle} show that it converges, as $t \to \infty$, to the discrete nonparametric MLE.  It would be interesting to determine a stopping criterion such that the corresponding estimator could be called a smooth nonparametric {\em near-MLE}.  Alternatively, one can pick any suitable density estimate $\hat f$ and plug in to \eqref{eq:fredholm}.  Numerical results indicate that this procedure will produce high-quality estimates of the mixing density, but its theoretical properties are still under investigation.

\section{Open problems}
\label{S:open}


\begin{problem}
This one goes all the way back to Newton's original development.  What is PR doing?  Is there any precise sense in which PR, or its permutation-averaged version, gives an approximation to the Dirichlet process mixture Bayes estimator?  \citet{nqz} showed that the connection is exact for $n=1$ and they also investigated the case of $n=2$.  In particular, for two observations, $Y_1$ and $Y_2$, they show that both PR and the Dirichlet process mixture posterior mean take the form 
\[ a_0 p_0(u) + a_1 p_0(u \mid Y_1) + a_2 p_0(u \mid Y_2) + a_{12} p_0(u \mid Y_1, Y_2), \]
where $p_0(u \mid Y_i) \propto k(Y_i \mid u) p_0(u)$ and 
\[ p_0(u \mid Y_1,Y_2) \propto k(Y_1 \mid u) k(Y_2 \mid u) p_0(u), \]
the only difference being in the coefficients $a_1, a_2, a_{12}$.  It can also be shown that the permutation-averaged PR estimator is of the same form, again with different coefficients, but I will not list these here.  For the general $n$ case, if one imagines averaging the PR expression in Proposition~12 of \citet{ghoshtokdar} over different permutations of the data sequence, one can vaguely see something reminiscent of the Dirichlet process mixture posterior mean expression given in \citet{lo1984}.  So it seems like something interesting could be there, but the details have eluded me so far.  
\end{problem}

\begin{problem}
Existing implementations of PR have used numerical integration to evaluate the normalizing constant at each iteration.  So even though the theory puts no restriction on the dimension of the latent variable space, the reliance on quadrature methods makes it difficult to handle mixture over more than one or two dimensions.  Is it possible to use Monte Carlo methods to compute this integral?  A strategy that works with a fixed set of particles with weights that are updated at each iteration seems particularly promising, but these weights would need to be monitored carefully.  
\end{problem}

\begin{problem}
A bound on the convergence rate of the PR estimator was stated in Theorem~\ref{thm:pr}, but I noted that this bound is conservative in the sense that it seems to be attained when the mixing distribution is a point mass, not a smooth density.  So a relevant question is how one could incorporate smoothness assumptions about the true density to improve upon this rate?  
\end{problem}

\begin{problem}
The PR algorithm is naturally sequential and would be ideal in cases where the data ordering matters, e.g., dependent data problems.  However, currently nothing is known about PR in such cases; in fact, even defining the PR algorithm in such cases is not clear.  A suggestion is made in \citet{ghosalroy2009} but, to my knowledge, no one has pursued this direction at all.  
\end{problem}

\begin{problem}
For the PR-based estimate of the structural parameter $\theta$ described in Section~\ref{S:extension}, currently very little is known about its theoretical properties.  I indicated there that simulation experiments suggest an asymptotic normality result holds, but this has yet to be rigorously demonstrated.  In classical iid problems, the log-likelihood is additive and the central limit theorem can be used after linearization.  For the PR likelihood, however, the $i^\text{th}$ term depends---in a complicated way---on all of $Y_1,\ldots,Y_i$.  Martingale laws of large numbers and central limit theorems seem promising but, unfortunately, no progress has been made along these lines yet. 
\end{problem}

\begin{problem}
I mentioned a few high-dimensional empirical Bayes applications here in this review and, for these problems, I always felt that there should, at least in some cases, be a theoretical benefit to plugging in a smooth estimate of the prior density compared to, say, a discrete estimate like in \citet{jiang.zhang.2009}.  Unfortunately, I have not yet been able to identify a theoretical benefit, but I still believe that one exists. 
\end{problem}


\section{Conclusion}
\label{S:discuss}

In this paper, I have reviewed the work on theory and applications of the PR algorithm for estimating mixing distributions, along the way highlighting some new developments and some open problems.  This is only one of the many areas that JKG had an impact so, naturally, my review here made connections to a number of adjacent topics on which JKG worked, including Bayesian nonparametrics, density estimation, and high-dimensional testing and estimation.  That these are still the ``hot topics'' in the statistics literature is surely no coincidence, it is an testament to JKG's incredible foresight and influence.  I was so tremendously lucky to have had the opportunity to know and to work with JKG, and it is an honor to dedicate this work to him.  


\section*{Acknowledgments}

This work is partially supported by the National Science Foundation, DMS--1737929.  


\bibliographystyle{apalike}
\bibliography{/Users/rgmarti3/Dropbox/Research/mybib}

\end{document}